\title{ \textbf{Reduction of Mathematical Models of Nuclear Receptor Binding to Promoter
Regions}}
\author{Sarbaz H. A. Khoshnaw  \\ Department of Mathematics, University of Leicester, LE1  7RH, UK \\ E-Mail: sarbazmath@yahoo.com  }
\documentclass[a4paper]{article}         
\usepackage{chemarr, graphicx, subfigure}
\usepackage{amsmath, amsthm, amssymb, amsfonts} 

\begin{document}  
 \maketitle 
\Large{\textbf{Abstract}}\\   
{\Large We study kinetic model of Nuclear Receptor Binding to Promoter Regions. This model is written as a system of ordinary differential equations. Model reduction techniques have been used to simplify chemical kinetics. In this case study, the technique of Pseudo-first order approximation is applied to simplify the reaction rates. CellDesigner has been used to draw the structures of chemical reactions of Nuclear Receptor Binding to Promoter Regions. After model reduction, the general analytical solution for reduced model is given and the number of species and reactions are reduced from 9 species and 6 reactions to 6 species and 5 reactions.\\

}
\section {\LARGE {Introduction } }
 { \Large The classical theory of chemical kinetics is used to show biological processes in terms of mathematical modeling. The assumption is that a model consists of:  
\begin{description} 
\item[$\bullet$] A set of components (species)  
 $S=(S_{1} , S_{2}, ... , S_{m})$,\\
 for each component $S_{i}$, $i=1,2,...,m$ a non negative variable $[S_{i}]$ (Concentration of $S_{i}$) is defined; the vector of concentrations is $[S]$.    
\item[$\bullet$] A set of reactions
 $V=(v_{1}, v_{2}, ... , v_{n}).$  
\item[$\bullet$] A set of kinetic constants 
$K=(k_{1}, k_{2}, ... , k_{n})$.    
\end{description}
\noindent For a general equation with $m$ species $S_{1}, . . . , S_{m}$ and associated stoichiometric coefficients (the non-negative integers) $\alpha_{s1}, . . . , \alpha _{sm}$ (reactants) and $\beta_{s1}, . . . ,\beta_{sm}$ (products),each elementary reaction is represented by its stoichiometric equation as follows:\\ 

$\sum\limits_{i=1}^m \alpha_{si} S_{i} { \overset{k}{\longrightarrow}}\sum\limits_{i=1}^m \beta_{si} S_{i},$ \\
\noindent where $s$ enumerates the elementary reaction. The corresponding reaction rate $v$ is given by\\
$v = k \prod\limits_{i=1}^m [S_{i}]^{\alpha_{si}}$,\\ 
\noindent where $k > 0$ is the reaction rate coefficient. It is important to note that the reaction rates depend on the reactants but not on the products \cite{Gorban2010,Tamas}. \\     
The stoichiometric matrix is $N=(\gamma_{si}),$ where $\gamma_{si}=\beta_{si}-\alpha_{si}, i=1,2,..,m$ . The stoichiometeric vector $\gamma_{s}$ is the \textit{s}th row of $N$ with coordinates $\gamma_{si}=\beta_{si}-\alpha_{si}$ \cite{Yablonskii}. \\  

\noindent The standard mass action formula is applied to find the reaction rates. The system of ODE describes the dynamics of chemical reactions. The kinetic equations are:
\begin{equation}
\begin{array}{llll}  
\dfrac{d[S](t)}{dt}=N   V([S], K, t),
[S](0)=[S]_{0},   t \in [0,T], T \in R^{+}, 
\end{array}
\label{eq1}
 \end{equation}

\noindent where $N$ is a stoichometric matrix of $m$ by $n$, $[S](0)$ is an initial value of concentrations. \\

\noindent The differential equation for a particular component $(A)$ in a model is written as:\\
$ \dfrac{d[A]}{dt}=\sum v_{A,produced}-\sum v_{A, consumed} $\\
\noindent this means ( Rate of change of component $(A)$)=(Amount of $(A)$ formed in all reactions)-(Amount of $(A)$ consumed in all reactions), where $v_{A}$ is the rate of formation/consumption of species $A$ in a particular reaction \cite{Singh}.\\
\noindent A simple example of linear reactions is given as follows:
\begin{equation}
\begin{array}{llll}  
S_{1}{ \overset{k_{1}}{\longrightarrow}} S_{2}{ \overset{k_{2}}{\longrightarrow}}S_{3}{ \overset{k_{3}}{\longrightarrow}}S_{4}{ \overset{k_{4}}{\longrightarrow}}S_{5}
\end{array}\label{eqy}
\end{equation}
\noindent where the reaction rates of equation \ref{eqy} are defined,\\
$v_{1}=k_{1}[S_{1}], v_{2}=k_{2}[S_{2}], v_{3}=k_{3}[S_{3}], v_{4}=k_{4}[S_{4}].$ \\
The system of ODE for the above linear reactions is given as a matrix equation of the form:\\
\begin{equation}
\begin{array}{llll}  
\dfrac{d[S]}{dt}=N V
\end{array}\label{eqy1}  
\end{equation}
where \\$[S]=\begin{pmatrix}
[S_{1}] \\
[S_{2}] \\
[S_{3}] \\  
[S_{4}] \\
[S_{5}] \\
\end{pmatrix}$
,$N=\begin{pmatrix}
-1 & 0 & 0 & 0  \\
1 & -1 & 0 & 0 \\
0 & 1 & -1 & 0\\
0 & 0 & 1 & -1 \\
0 & 0 & 0 & 1 \\
\end{pmatrix}$, $V=\begin{pmatrix}
v_{1} \\
v_{2} \\
v_{3} \\
v_{4} 
\end{pmatrix}$ 
\subsection{\LARGE{Order and Molecularity}}
\textbf{\textit{Rate Constant}}: For a general reaction\\ 
$aA+bB{ \overset{k}{\longrightarrow}} cC+dD$\\
the rate is given as follows:\\
Rate$=k[A]^{a}[B]^{b} $\\  
where $k$ is a rate constant or velocity constant \cite{Upadhyay,Malijevsky}. The rate constant for any reaction can be found either by measuring the reaction rate at unit concentrations for the reactants or by knowing the rate of the reaction using the following relation:\\
Rate constant$=\dfrac{Rate}{[A]^{a}[B]^{b}}.$\\
\textbf{\textit{Order and Molecularity}}:For a giving reaction\\
$aA+bB+cC+...{ \overset{k}{\longrightarrow}} product$\\
the reaction rate is defined by:\\
Rate$=k[A]^{a}[B]^{b}[C]^{c}... $\\
Then the reaction has $a-th$ order with respect to $A$, $b-th$ order with respect to $B$, $c-th$ order with respect to $C$, ..., and the overall order of reaction is $a+b+c+...=n.$\\
The summation of stoichiometric coefficients of reactions is called \textbf{\textit{Molecularity}} of reaction. For instance, for a giving stoichiometric equation: $3A+2B=C+4D$, the stoichiometric coefficients of $A$ and $B$ are 3 and 2, respectively \cite{Upadhyay}. Therefore, the molecularity of the reaction is $3+2=5$.
The \textbf{\textit{rate of complex reactions}} (multi-step reactions) is determined by the rate of change of product (the rate of increase of product). The order and molecularity of a reaction have not a simple relationship. If we have a reaction which occurs in two or more different steps, and it gives overall the same reaction then the order and molecularity of reaction are different. Let give some examples to differentiate between molecularity and order of reaction.\\
For example, consider a reaction in two steps:
\begin{center}
Step 1: Fast  $I+B { \overset{k_{1}}{\longrightarrow}} P$\\
Step 2: Slow $A+2B{ \overset{k_{2}}\longrightarrow}I$\\
 \line(1,0){120} \\
$A+3B{ \overset{k}\longrightarrow}P$
\end{center}
\noindent Note that, the rate of complex reactions (Multi-step reactions) will depend on the rate of increase of product. Therefore, the rate of product is expected as follows\\
\begin{equation}
\begin{array}{llll}  
Rate\approx \dfrac{d[P]}{dt}=k_{1}[I][B]
\end{array}\label{eqh1}  
\end{equation}
\noindent and the rate of intermediate species $I$ can be given:
\begin{equation}
\begin{array}{llll}  
\dfrac{d[I]}{dt}=k_{2}[A][B]^{2}-k_{1}[I][B]
\end{array}\label{eqh2}  
\end{equation}
\noindent the steady state with respect to $\dfrac{d[I]}{dt}$ can be applied because it is usually present in very small concentrations. This means $\dfrac{d[I]}{dt}\approx0$. Thus, from equation \ref{eqh2} , we get $[I]=\dfrac{k_{2}}{k_{1}}[A][B]$, and put the value of $[I]$ in equation \ref{eqh1}, we obtain the rate of the complex reaction:
\begin{equation}
\begin{array}{llll}  
Rate\approx \dfrac{d[P]}{dt}=k_{2}[A][B]^{2}
\end{array}\label{eqh3}  
\end{equation}
\noindent As a result, the order of reaction would be one with respect to reactant $A$ and two with respect to reactant $B$, and the overall order of the reaction is 3. On the other hand, the molecularity of the above reaction would be $1+3=4.$ Thus, the order and molecularity of the reaction are not the same. \\
\noindent Another example is the reaction of cane sugar:
\begin{center}
$C_{12}H_{22}O_{11}+H_{2}O{ \overset{k}{\longrightarrow}} C_{6}H_{12}O_{6}+C_{6}H_{12}O_{6}$
\end{center}
the reaction rate is given as: 
\begin{equation}
\begin{array}{llll}  
Rate=k[Sucrose][H_{2}O]
\end{array}\label{s1}
\end{equation}
\noindent This reaction looks like second order, first order with respect to any reactant $Sucrose$ and $H_{2}O$  \cite{Upadhyay}. The $[H_{2}O]$ remains constant, and it is present in large excess. Therefore, the reaction is only first order with respect to $Sucrose$ because the $[H_{2}O]$ does not effect the rate of reaction. As a result, the equation \ref{s1} would be written as:
\begin{center}
Rate$=k^{*}[Sucrose]$
\end{center} 
where $k^{*}=k[H_{2}O]_{0}$, and $[H_{2}O]_{0}$ is the initial value of $[H_{2}O]$. This reaction is called \textbf{\textit{Pseudo-first order reaction}}. If the order and molecularity of a reaction are different as a result of one reactant in excess, then the reaction is known as \textbf{\textit{Pseudo- first order reaction.}}\\
\noindent However, if we have a reaction which occurs in a single step then the order of reaction and molecularity are the same. Consider the reaction:
$2A+4B{ \overset{k}{\longrightarrow}} P$\\
the molecularity of the reaction is $2+4=6$. If the above reaction occurs just in a single step, then the overall order of the reaction is 6 because the reaction rate is: Rate$=k[A]^{2}[B]^{4}$. Thus, the order and molecularity in this reaction are the same value.
\section{\LARGE{Model Reduction}}
\noindent Transformation of the system (\ref{eq1}) to another system is called"\textbf{model reduction}" in which the new system includes smaller number of equations without affecting dynamics of variables $ [S_{1}](t), [S_{2}](t), ... , [S_{m}](t)$ \cite{Radulescu, Choi, Conzelmann}. We apply some techniques of reduction to the system, reducing it to essentially fewer species and reactions. There are some model reduction techniques which have been used to simplified the model. \\

\textbf{\LARGE{Methods}}
\subsection{\LARGE{Pseudo-First Order Approximation}}
In this technique the order of reaction is smaller than the actual reaction order. This is sometimes happen for a reaction with two or more reactants. If one reactant is present in large excess, its concentration change is negligible small \cite{Upadhyay,Malijevsky}. Therefore, the reaction rate becomes independent from this reaction. Consider a reaction 
\begin{center}
$A+B{ \overset{k}{\longrightarrow}}P$
\end{center}
with reaction rate:
\begin{equation}
\begin{array}{llll}  
v=k[A][B]
\end{array}\label{s2}
\end{equation}
This reaction is first order with respect to $A$ and first order with respect to $B$, and overall is second order. If the initial of $[B]$ is present in large amount, and the initial value of $[A]$ is smaller, then the effect of $A$ on the concentration and reaction rate is much greater than the effect of $B$. To explain this idea, let give some numerical values: \\
Let $ [A]_{0}=4 $ and $ [B]_{0}=100 $ at the time where half of $ A $ has consumed away: $ [A]=2 $ and $ [B]=98 $. Thus, $ [A] $ has changed by $0.50$ while $ [B] $ changed only by $0.02$. By comparing reaction rate with initial rate we obtain: \\
$ v=k (0.50 [A]_{0}) (0.98 [B]_{0}) =0.49[A]_{0}[B]_{0}=0.49 v_{0}. $\\
It is clear that the decreasing in reaction rate is almost determined by the change of $[A]$, because the relative change of concentration of $A$ is much larger than the relative change of $B$.\\
As a result, the change of $[B]$ can be ignored throughout the reaction proceeds. In this case the rate of reaction is not effected by $[B]$, and the reaction is simply first order with respect to $A$. In other words, if $[B]$ is present in large excess (i.e $[B](t)\gg[A](t)$ $\forall$ $t\in [0,T]$), and $[B]$ remains constant (i.e $[B]\approx[B]_{0}$, and $[B]$ does not significantly change), then the rate equation \ref{s2} can be written as a linear equation:
\begin{center}
$v=k^{*}[A]$
\end{center}
where $k^{*}=k[B]_{0}$. Thus, the reaction is called pseudo-first order approximation. \\
\noindent \textbf{Similarly}, if we have a third order reaction
\begin{center}
$A+B+C{ \overset{k_{1}}{\longrightarrow}}P$
\end{center}
with reaction rate:
\begin{equation}
\begin{array}{llll}  
v=k_{1}[A][B][C]
\end{array}\label{s3}
\end{equation}
and two of the reactants are present in large excess, let $[A]$ and $[B]$ are in large amount. This means $[A](t)\gg[C](t)$ and $[B](t)\gg[C](t)$ $\forall$ $t\in [0,T]$. In this case $[A]$ and $[B]$ do not significantly affect on the reaction rate (Equation \ref{s3}), and they remain constant ($[A]\approx[A]_{0}$ and $[B]\approx[B]_{0}$). Therefore, the rate equation \ref{s3} would be considered as a linear equation: 
\begin{center}
$v=k_{1}^{*}[C]$, where $k_{1}^{*}=k_{1}[A]_{0}[B]_{0}$
\end{center}
As a result, the reaction is called pseudo-first order approximation. \\
\noindent \textbf{Generally}, consider \textit{n}th order reaction
\begin{center}
$A_{1}+A_{2}+...+A_{n}{ \overset{k_{2}}{\longrightarrow}}P$
\end{center}
with reaction rate:
\begin{equation}
\begin{array}{llll}  
v = k_{2} \prod\limits_{i=1}^n [A_{i}],
\end{array}\label{s41}
\end{equation}
and $n-1$ reactants are present in large excess.\\ Let $[A_{1}], [A_{2}], ..., [A_{n-1}]$ are in large amount. This means $[A_{i}](t)\gg[A]_{n}(t) , i=1,2,...,n-1,$ $\forall t\in[0,T]$. In this case $[A_{1}], [A_{2}], ..., [A_{n-1}]$ do not significantly affect on the reaction rate (Equation \ref{s41}), and they remain constant ($[A_{i}]\approx[A_{i}]_{0},$ for $i=1,2,...,n-1$). Therefor, the rate equation \ref{s41} would be considered as a linear equation: 
\begin{center}
$v=k_{2}^{*}[A]_{n}$, where $k_{2}^{*}=k_{2} \prod\limits_{i=1}^{n-1} [A_{i}]_{0}$
\end{center}
Thus, the reaction is called pseudo-first order approximation. 
\subsection{\LARGE{Removal of approximately linearly dependent concentrations}}
This assumption is that if we have two species $B$ and $C$ in a model, $[B]$ and $[C]$ are approximately linearly dependent (i.e $[B](t)\approx k[C](t), \forall t \in [0,T], k \in R$ ), then one of them can be neglected from the model \cite{Kutumova}. For example, for the parallel reactions \cite{Malijevsky}:  
 \begin{equation}
\begin{array}{llll}  
A{ \overset{k_{1}}{\longrightarrow}}B\\
A{ \overset{k_{2}}{\longrightarrow}}C
\end{array}\label{s4}
\end{equation}
\noindent and the system of ODE is given: 
 \begin{equation}
\begin{array}{llll}  
\dfrac{d[A]}{dt}=-(k_{1}+k_{2})[A],
\end{array}\label{s5}
\end{equation}
 \begin{equation}
\begin{array}{llll}  
\dfrac{d[B]}{dt}=k_{1}[A],
\end{array}\label{s6}
\end{equation}
\begin{equation}
\begin{array}{llll}  
\dfrac{d[C]}{dt}=k_{2}[A],
\end{array}\label{s7}
\end{equation}
\noindent with the initial concentrations:
\begin{equation}
\begin{array}{llll}  
[A](0)=[A]_{0}, [B](0)=0, [C](0)=0.
\end{array}\label{s8}
\end{equation}
\noindent The analytical solution for equation \ref{s5} with initial conditions(Equation \ref{s8}) is given as:
\begin{equation}
\begin{array}{llll}  
[A](t)=[A]_{0} e^{-(k_{1}+k_{2})t}
\end{array}\label{s9}
\end{equation} 
\noindent Substitution of this equation for $[A]$ into the equations \ref{s6} and \ref{s7}, then by straightforward integration, we obtain:
\begin{equation}
\begin{array}{llll}  
[B](t)=\dfrac{k_{1}[A]_{0}}{k_{1}+k_{2}} (1-e^{-(k_{1}+k_{2})t})
\end{array}\label{s10}
\end{equation} 
and
\begin{equation}
\begin{array}{llll}  
[C](t)=\dfrac{k_{2}[A]_{0}}{k_{1}+k_{2}} (1-e^{-(k_{1}+k_{2})t})
\end{array}\label{s11}
\end{equation} 
\noindent Dividing equation \ref{s10} by equation \ref{s11}, we get $[B]=\frac{k_{1}}{k_{2}}[C]$, this means that $[B]$ and $[C]$ are linearly dependent. Thus, either $[B]$ or $[C]$ can be neglected from the model because they have the same chemical kinetic properties.
\subsection{\LARGE{Removal of approximately linearly dependent reaction rates}}
This technique is used to neglect the linearly dependent reaction rates. If we have two reaction rates $v_{1}$ and $v_{2}$ in a model, and they are linearly dependent (i.e $v_{1} \approx kv_{2}, k \in R$), then either $v_{1}$ or $v_{2}$ can be neglected from the model. To explain this idea, let give the following parallel reactions:
\begin{equation}
\begin{array}{llll}  
A+B{ \overset{k_{1}}{\longrightarrow}}C\\
A+B{ \overset{k_{2}}{\longrightarrow}}D
\end{array}\label{s12}
\end{equation}  
\noindent with reaction rates:
\begin{equation}
\begin{array}{llll}  
v_{1}=k_{1}[A][B],
\end{array}\label{s13}
\end{equation}  
\begin{equation}
\begin{array}{llll}  
v_{2}=k_{2}[A][B].
\end{array}\label{s14}
\end{equation}  
\noindent Dividing equation \ref{s13} by equation \ref{s14}, we obtain:\\
$\dfrac{v_{1}}{v_{2}}=\dfrac{k_{1}[A][B]}{k_{2}[A][B]} \Longrightarrow v_{1}=\dfrac{k_{1}}{k_{2}} v_{2}.$
\noindent Thus, $v_{1}$ and $v_{2}$ are linearly dependent, and one of them would be neglected from the model.  
 \section {\LARGE {Nuclear Receptor Binding to Promoter Regions}} 
Nuclear receptor Binding to Promoter regions consists of 9 species and 6 reactions (See figure \ref{nrb1} and table \ref{table1}). This model was presented in previous study \cite{Kolodkin}. In this work, we suggested a mathematical model for the chemical reactions, and we then simplified the model by using the technique of pseudo-first order approximation. The reaction rates in this work are based on the stander mass action kinetics. It means that mass action formula is used to find the reaction rates. The reactions of the complete model and reduced model are presented in tables \ref{table1} and \ref{table2}, respectively. Celldesigner has been used to simulate the concentration of the species. 
  \subsection{\LARGE{Mathematical Modeling of the Nuclear receptor Binding to Promoter regions}}
The chemical network reactions of this model ( Figure \ref{nrb1}) can be written as a system of ordinary differential equations. This means that the mass action formula is applied to find the reaction rates for chemical kinetics ( Table \ref{table1}). Thus, this model is given by the following system of differential equations:\\
 \begin{equation}
\begin{array}{llll}
\dfrac{d[hsp90:NR]}{dt}=-v_{20}-v_{21}-v_{22}, 
\end{array}\label{eq23}
\end{equation}
\begin{equation}
\begin{array}{llll}
 \dfrac{d[Re]}{dt}=-v_{18}-v_{20},
\end{array}\label{eq24}
\end{equation}
\begin{equation}
\begin{array}{llll}
\dfrac{d[L:2NR]}{dt}=-v_{18}-v_{19},
\end{array}\label{eq25}
\end{equation}
\begin{equation}
\begin{array}{llll}
\dfrac{d[hsp90:NR:Re]}{dt}=v_{20}-v_{21}-v_{23},
\end{array}\label{eq26}
\end{equation}
\begin{equation}
\begin{array}{llll}
\dfrac{d[hsp90:2NR:Re]}{dt}=v_{21}-v_{22},
\end{array}\label{eq27}
\end{equation}
\begin{equation}
\begin{array}{llll}
\dfrac{d[hsp90:3NR:Re]}{dt}=v_{22}-v_{23},  
\end{array}\label{eq28}  
\end{equation}
\begin{equation}
\begin{array}{llll}
\dfrac{d[hsp90:4NR:Re]}{dt}=v_{23},
\end{array}\label{eq28}
\end{equation}
\begin{equation}
\begin{array}{llll}
 \dfrac{d[L:2NR:Re]}{dt}=v_{18}-v_{19},
\end{array}\label{eq30}
\end{equation}
\begin{equation}
\begin{array}{llll}
\dfrac{d[L:4NR:Re]}{dt}=v_{19},
\end{array}\label{eq31}
\end{equation}
\noindent where the reaction rates of the above equations are defined as follows:\\
\noindent $ v_{18}=k_{18} [L:2NR] [Re], $ \\ 
\noindent $ v_{19}=k_{19} [L:2NR:Re] [L:2NR],  $ \\ 
\noindent $ v_{20}=k_{20} [hsp90:NR] [Re],$ \\ 
\noindent $ v_{21}=k_{21}[hsp90:NR:Re] [hsp90:NR], $ \\  
\noindent $ v_{22}=k_{22}[hsp90:2NR:Re] [hsp90:NR],  $ \\ 
\noindent $ v_{23}=k_{23}[hsp90:NR:Re] [hsp90:3NR:Re]. $ \\ 
\noindent The initial value of concentrations are given as follows: \\
\noindent $ [hsp90:NR](0)=[hsp90:NR]_{0} , [Re](0)=[Re]_{0}, $ \\
\noindent $  [hsp90:NR:Re](0)=0, [hsp90:2NR:Re](0)=0,  $ \\
\noindent $  [hsp90:3NR:Re](0)=0, [hsp90:4NR:Re](0)=0,  $ \\
\noindent $  [L:2NR](0)=[L:2NR]_{0}, [L:2NR:Re](0)=0,  $ \\
\noindent $ [L:4NR:Re](0)=0.  $ \\
\noindent The system of differential equations (Equations \ref{eq23}-\ref{eq31}) can be written as a matrix equation of the form:
\begin{equation}
\begin{array}{llll}  
\dfrac{d[S]}{dt}=N V
\end{array}\label{eqy2}
\end{equation} 
where \\ 
\begin{gather*}
  \setlength{\arraycolsep}{1\arraycolsep}
  \text{\Large$ 
[S] =\begin{pmatrix}
[hsp90:NR] \\
[Re] \\ 
[L:2NR] \\
[hsp90:NR:Re] \\
[hsp90:2NR:Re] \\
[hsp90:3NR:Re] \\
[hsp90:4NR:Re] \\
[L:2NR:Re] \\
[L:4NR:Re] \\
\end{pmatrix}
, V =\begin{pmatrix}
v_{18} \\
v_{19} \\
v_{20} \\
v_{21} \\
v_{22} \\
v_{23} \\
\end{pmatrix}
$}
\end{gather*} 
and 
\begin{gather*}
  \setlength{\arraycolsep}{1\arraycolsep}
  \text{\Large $ 
  N=  \begin{pmatrix}
       0 & 0 & -1 & -1& -1 & 0 \\
       -1 & 0 & -1 & 0& 0 & 0 \\
       -1 & -1 & 0 & 0& 0 & 0 \\
       0 & 0 & 1 & -1& 0 & -1 \\
       0 & 0 & 0 & 1& -1 & 0 \\
       0 & 0 & 0 & 0& 1 & -1 \\
       0 & 0 & 0 & 0& 0 & 1 \\
       1& -1 & 0 & 0& 0 & 0 \\
       0 & 1 & 0 & 0& 0 & 0 \\    
    \end{pmatrix}
    $}
\end{gather*}
\section{\LARGE{Results}}   
In this work, CellDesigner has been used to draw the structures of chemical reactions of Nuclear Receptor Binding to Promoter Regions (NRB) (Figures \ref{nrb1} and \ref{nrb2}). We use the value of rate constants (Table \ref{table3}) and initial concentrations (Table \ref{table4}) to simulate of concentrations. The simplification of reaction rates is mainly based on the technique of pseudo-first order approximation. We presented the results of this work as follows:
\begin {enumerate} 
\item \textbf{Simplification of kinetic equations based on pseudo-first order approximation}: It is noticed that the concentration of $hsp90:NR$ over all is greater than the concentration of $hsp90:NR:Re$ and $hsp90:2NR:Re$, respectively (See figure \ref{re1}), and the concentration of $L:2NR$ is also grater then the concentration of $L:2NR:Re$ (See figure \ref{re2}). This means:\\
\noindent $[hsp90:NR](t) \gg [hsp90:NR:Re](t),$\\
\noindent  $ [hsp90:NR](t) \gg [hsp90:2NR:Re](t),$\\
\noindent  $[L:2NR](t) \gg [L:2NR:Re](t),$ $\forall$ $t \in [0,T], T \in R^{+}$.\\
\noindent In other words, $[hsp90:NR]$ and $[L:2NR]$ are in large excess. They remain relatively constant (i.e $[hsp90:NR] \approx[hsp90:NR]_{0}$ and $[L:2NR]\approx[L:2NR]_{0}$). Therefore, the reaction rates 19, 21 and 22 are changed as follows:\\
\noindent $v_{19}^{*} =k_{19}^{*} [L:2NR:Re],$  \\
\noindent $v_{21}^{*} =k_{21}^{*} [hsp90:NR:Re],$  \\
\noindent $v_{22}^{*} =k_{22}^{*} [hsp90:2NR:Re],$ \\
\noindent where $k_{19}^{*} =k_{19} [L:2NR]_{0}, k_{21}^{*}=k_{21} [hsp90:NR]_{0}$ and $k_{22}^{*}=k_{22} [hsp90:NR]_{0}$.\\  
\noindent In addition, the concentration of $Re$ is dominated by the concentration of $hsp90:NR$ and $L:2NR$ ( Figure \ref{re1} and \ref{re2}), it means $[hsp90:NR](t) \gg [Re](t)$, and $[L:2NR](t) \gg [RE](t)$, $\forall$ $t\in [0,T]$. They are in large amount, and they remain approximately unchanged.  According to the technique of pseudo-first order approximation, the reaction rates 18 and 20 can be simplified as follows\\
\noindent $ v_{18}^{*}=k_{18}^{*}[Re], $ and $ v_{20}^{*}=k_{20}^{*}[Re], $\\
where,\\  
\noindent $k_{18}^{*}=k_{18} [hsp90:NR]_{0}$ and $ k_{20}^{*}=k_{20} [L:2NR]_{0} $.
\item \textbf{Removal of slow reaction}:The reaction 23 in this model is the slowest reaction in comparison with other reactions in the model. This reaction can be ignored from the model because it does not significantly affect on the experimental dynamic of chemical reactions.
\end{enumerate}
\section{\LARGE{Reduced model of Nuclear Receptor Binding to Promoter Regions}}
The model of nuclear receptor binding to promoter regions is reduced after applying the above techniques of reduction. The simplified model includes 6 species and 5 reactions (Figure \ref{nrb2} and table \ref{table2}). It can be presented as a system of ODE. This is given as follows:
\begin{equation}
\begin{array}{llll}
\dfrac{d[Re]}{dt}=-v_{18}^{*}-v_{20}^{*}, 
\end{array}\label{eq32}
\end{equation}
\begin{equation}
\begin{array}{llll}
 \dfrac{d[L:2NR:Re]}{dt}=v_{18}^{*}-v_{19}^{*},
\end{array}\label{eq33}
\end{equation}
\begin{equation}
\begin{array}{llll}
\dfrac{d[L:4NR]}{dt}=v_{19}^{*},
\end{array}\label{eq34}
\end{equation}
\begin{equation}
\begin{array}{llll}
\dfrac{d[hsp90:NR:Re]}{dt}=v_{20}^{*}-v_{21}^{*},
\end{array}\label{eq35}
\end{equation}
\begin{equation}
\begin{array}{llll}
\dfrac{d[hsp90:2NR:Re]}{dt}=v_{21}^{*}-v_{22}^{*},
\end{array}\label{eq36}
\end{equation} 
\begin{equation}
\begin{array}{llll}
\dfrac{d[hsp90:3NR:Re]}{dt}=v_{22}^{*},
\end{array}\label{eq37}
\end{equation}  
\noindent The system of differential equations (Equations \ref{eq32}-\ref{eq37}) can be written as a matrix equation of the form:
\begin{equation}
\begin{array}{llll}  
\dfrac{d[S]^{*}}{dt}=N^{*} V^{*}
\end{array}\label{eq38}
\end{equation}
where \\ 
\begin{gather*}
  \setlength{\arraycolsep}{1\arraycolsep}
  \text{\Large$ 
[S]^{*} =\begin{pmatrix}
[Re] \\
[L:2NR:Re] \\
[L:4NR:Re] \\ 
[hsp90:NR:Re] \\
[hsp90:2NR:Re] \\
[hsp90:3NR:Re] \\
\end{pmatrix}
,  V^{*} =\begin{pmatrix}
v_{18}^{*} \\
v_{19}^{*} \\
v_{20}^{*} \\
v_{21}^{*} \\
v_{22}^{*} \\
\end{pmatrix}
$}
\end{gather*} 
and 
\begin{gather*}
  \setlength{\arraycolsep}{1\arraycolsep}
  \text{\Large $ 
  N^{*}=  \begin{pmatrix}
 -1 & 0 & -1 & 0 & 0 \\
 1 & -1 & 0 & 0 & 0 \\
 0 & 1& 0 & 0 & 0 \\
 0 & 0 & 1 & -1 & 0 \\
 0 & 0 & 0 & 1 & -1 \\
 0 & 0 & 0 & 0 & 1 \\
    \end{pmatrix}
  $}
\end{gather*}  
\noindent The general analytical solution of the system \ref{eq38} is given as follows:\\
\noindent $\textbf{[Re](t)}= -c_{4}(3k_{18}^{*}(k_{20}^{*})^2 + 3(k_{18}^{*})^2 k_{20}^{*} - (k_{18}^{*})^2 k_{21}^{*} \\
- (k_{18}^{*})^2 k_{22}^{*} - (k_{20}^{*})^2 k_{21}^{*} - (k_{20}^{*})^2 k_{22}^{*} + (k_{18}^{*})^3 + (k_{20}^{*})^3 \\
- 2 k_{18}^{*} k_{20}^{*} k_{21}^{*}- 2 k_{18}^{*} k_{20}^{*} k_{22}^{*} + k_{18}^{*} k_{21}^{*} k_{22}^{*} + k_{20}^{*} k_{21}^{*} k_{22}^{*})e^{-(k_{18}^{*} + k_{20}^{*})t}\\
/(k_{20}^{*} k_{21}^{*} k_{22}^{*})$,\\
\noindent $\textbf{[L:2NR:Re](t)}=-c_{4} ((k_{18}^{*})^3 k_{21}^{*} - k_{18}^{*} (k_{20}^{*})^3 - 3 (k_{18}^{*})^3 k_{20}^{*} \\
-  (k_{18}^{*})^4 + (k_{18}^{*})^3 k_{22}^{*} - 3(k_{18}^{*})^2 (k_{20}^{*})^2 +k_{18}^{*} (k_{20}^{*})^2 k_{21}^{*} \\ + 2 (k_{18}^{*})^2 k_{20}^{*} k_{21}^{*}- k_{18}^{*} k_{20}^{*} k_{21}^{*} k_{22}^{*}+k_{18}^{*} (k_{20}^{*})^2 k_{22}^{*} \\
+ 2(k_{18}^{*})^2 k_{20}^{*} k_{22}^{*} - (k_{18}^{*})^2 k_{21}^{*} k_{22}^{*}) e^{-(k_{18}^{*} + k_{20}^{*})t}  \\
- c_{2}( k_{18}^{*}k_{20}^{*} k_{21}^{*} k_{22}^{*}  - k_{19}^{*} k_{20}^{*} k_{21}^{*} k_{22}^{*}+ (k_{20}^{*})^2 k_{21}^{*} k_{22}^{*}) e^{-k_{19}^{*}t}\\
/ ( (k_{20}^{*})^2 k_{21}^{*} k_{22}^{*}+ k_{18}^{*}k_{20}^{*}k_{21}^{*}k_{22}^{*} - k_{19}^{*}k_{20}^{*}k_{21}^{*}k_{22}^{*})$,\\
\noindent $\textbf{[L:4NR:Re](t)}=(c_{2}((k_{20}^{*})^2 k_{21}^{*} k_{22}^{*} + k_{18}^{*}k_{20}^{*}k_{21}^{*}k_{22}^{*}\\
- k_{19}^{*}k_{20}^{*}k_{21}^{*}k_{22}^{*})e^{-k_{19}^{*}t}-c_{4}((k_{18}^{*})^3 k_{19}^{*} +k_{18}^{*}k_{19}^{*}(k_{20}^{*})^2\\
+ 2(k_{18}^{*})^2 k_{19}^{*}k_{20}^{*}-(k_{18}^{*})^2 k_{19}^{*}k_{21}^{*}-(k_{18}^{*})^2 k_{19}^{*}k_{22}^{*}-k_{18}^{*}k_{19}^{*}k_{20}^{*}k_{21}^{*}\\
- k_{18}^{*}k_{19}^{*}k_{20}^{*}k_{22}^{*}+k_{18}^{*}k_{19}^{*}k_{21}^{*}k_{22}^{*})e^{-(k_{18}^{*}+k_{20}^{*})t}+ c_{5}((k_{20}^{*})^2 k_{21}^{*}k_{22}^{*}\\
+ k_{18}^{*}k_{20}^{*}k_{21}^{*}k_{22}^{*}-k_{19}^{*}k_{20}^{*}k_{21}^{*}k_{22}^{*}))/((k_{20}^{*})^2 k_{21}^{*}k_{22}^{*}+ k_{18}^{*}k_{20}^{*}k_{21}^{*}k_{22}^{*} \\
- k_{19}^{*}k_{20}^{*}k_{21}^{*}k_{22}^{*})$,\\
\noindent $\textbf{[hsp90:NR:Re](t)}=(c_{4}((k_{18}^{*})^2+ (k_{20}^{*})^2+ 2k_{18}^{*}k_{20}^{*}\\
 - k_{18}^{*}k_{22}^{*}- k_{20}^{*}k_{22}^{*}) e^{-(k_{18}^{*}+k_{20}^{*})t}+c_{1}((k_{21}^{*})^2- k_{21}^{*} k_{22}^{*}) e^{-k_{21}^{*} t})\\
/(k_{21}^{*}k_{22}^{*})$,\\
\noindent $\textbf{[hsp90:2NR:Re](t)}=-(c_{1} k_{21}^{*} e^{-k_{21}^{*}t} + c_{3}k_{22}^{*} e^{-k_{22}^{*}t}\\
+c_{4}k_{18}^{*} e^{-(k_{18}^{*}+k_{20}^{*})t}+ c_{4}k_{20}^{*}e^{-(k_{18}^{*}+k_{20}^{*})t})/k_{22}^{*}$,\\
\noindent $\textbf{[hsp90:3NR:Re](t)}=c_{1} e^{-k21k_{21}^{*}*t} + c_{3} e^{-k_{22}^{*}*t} \\
+ c_{4} e^{-(k_{18}^{*}+k_{20}^{*})t}+c_{6}$, \\
\noindent where $c_{1}, c_{2}, c_{3}, c_{4}, c_{5}$ and $c_{6}$ are constants.   
\section{Conclusion}
Mathematical modeling gives a powerful tool for investigation the properties of chemical kinetics. Methods of model reduction allowed us to reduce the number of reactions and species in the model of nuclear receptor binding to promoter regions.This model is presented as a system of ODEs. The methods of model reduction provide not only faster computational time, but they have a good benefit  to make the system simpler, making it easier to understand and manipulate. The technique of pseudo-first order approximation is used to simplify some kinetic equations when one reactant dominates others. We use Celldesigner to draw the network of chemical reactions and to simulate the concentration of species. After model reduction, the number of species and reactions are reduced from 9 species and 6 reactions to 6 species and 5 reactions. It could be said that the results in this work may have a real advantage of chemical kinetics of nuclear receptor binding to promoter regions.
 
\newpage
 \begin{table} 
  \center
  \begin{tabular}{ | p{0.5 cm} | l | l |   }
    \hline
    \textbf{No}& \Large \textbf{{Reactions(Kinetics) of Original Model of NRB}}  \\ \hline
    $k_{18}$ & $L:2NR+Re {\longrightarrow} L:2NR:Re$  \\ \hline
    $ k_{19} $ & $L:2NR+L:2NR:Re {\longrightarrow} L:4NR:Re$ \\ \hline   
    $ k_{20} $ & $hsp90:NR+Re {\longrightarrow} hsp90:NR:Re $ \\ \hline 
    $ k_{21} $ & $hsp90:NR+hsp90:NR:Re  {\longrightarrow} hsp90:2NR:Re $ \\ \hline    
    $ k_{22} $ & $hsp90:NR+hsp90:2NR:Re  {\longrightarrow}  hsp90:3NR:Re $ \\ \hline
    $ k_{23} $ & $hsp90:NR:Re+hsp90:3NR:Re  {\longrightarrow} hsp90:4NR:Re $ \\ \hline
        \end{tabular}
    \caption{ This table shows the reactions of the original model of Nuclear Receptor Binding to Promoter Regions. }
\label{table1} 
\end{table}  
      
  \begin{table} 
  \center
  \begin{tabular}{ | p{0.5 cm} | l | l |   }
    \hline
   \textbf{No}& \Large \textbf{{Reactions(Kinetics) of Reduced Model of NRB}}  \\ \hline
    $k_{18}^{*}$ & $Re {\longrightarrow} L:2NR:Re$  \\ \hline
    $ k_{19}^{*} $ & $L:2NR:Re {\longrightarrow} L:4NR:Re$ \\ \hline   
    $ k_{20}^{*} $ & $Re {\longrightarrow} hsp90:NR:Re $ \\ \hline 
    $ k_{21}^{*} $ & $hsp90:NR:Re  {\longrightarrow} hsp90:2NR:Re $ \\ \hline    
    $ k_{22}^{*} $ & $hsp90:2NR:Re  {\longrightarrow} hsp90:3NR:Re $ \\ \hline
        \end{tabular}
    \caption{This table presents the reactions of the reduced model of Nuclear Receptor Binding to Promoter Regions. }
\label{table2}   
\end{table}  
 \begin{table} 
  \center
  \begin{tabular}{  | l | l | l | l |   }
    \hline
     \textbf {{Rate constants}} & \textbf{{Values}}   \\ \hline
    $k_{18}$ & $ 8 \times 10^{-2}$   \\ \hline
    $k_{19}$  &   $ 1.03 \times 10^{0}$  \\ \hline
    $k_{20}$ & $ 4 \times 10^{-2}$   \\ \hline
    $k_{21}$ & $ 4 \times 10^{-3}$   \\ \hline
    $k_{22}$ & $2 \times 10^{-3} $   \\ \hline
    $k_{23}$ & $ 9 \times 10^{-1}$   \\ \hline
  
    \end{tabular}
    \caption{This table shows the value of rate constants. }
\label{table3} 
\end{table}  
 \begin{table} 
  \center
  \begin{tabular}{  | l | l | l | l |   }
    \hline
     \textbf {{Initial Conditions}} & \textbf{{Values}}  \\ \hline
  $[Re]_{0} $ & $ 0.8$  \\ \hline
  $[hsp90:NR]_{0}$ & $ 1.4$  \\ \hline
  $[hsp90:NR:Re]_{0}$ & $ 0$  \\ \hline
  $[hsp90:2NR:Re]_{0}$ & $ 0$  \\ \hline
  $[L:4NR:Re]_{0}$ & $ 0$  \\ \hline
  $[L:2NR:Re]_{0}$ & $ 0$  \\ \hline
  $[hsp90:3NR:Re]_{0}$ & $ 0$  \\ \hline
  $[L:2NR]_{0}$ & $ 1.2 $  \\ \hline
  $[hsp90:4NR:Re]_{0}$ & $0$  \\ \hline
  
    \end{tabular}
    \caption{This table shows the value of initial concentrations. }
\label{table4} 
\end{table}  
\begin{figure}[h]  
\center
\includegraphics[width=1\textwidth]{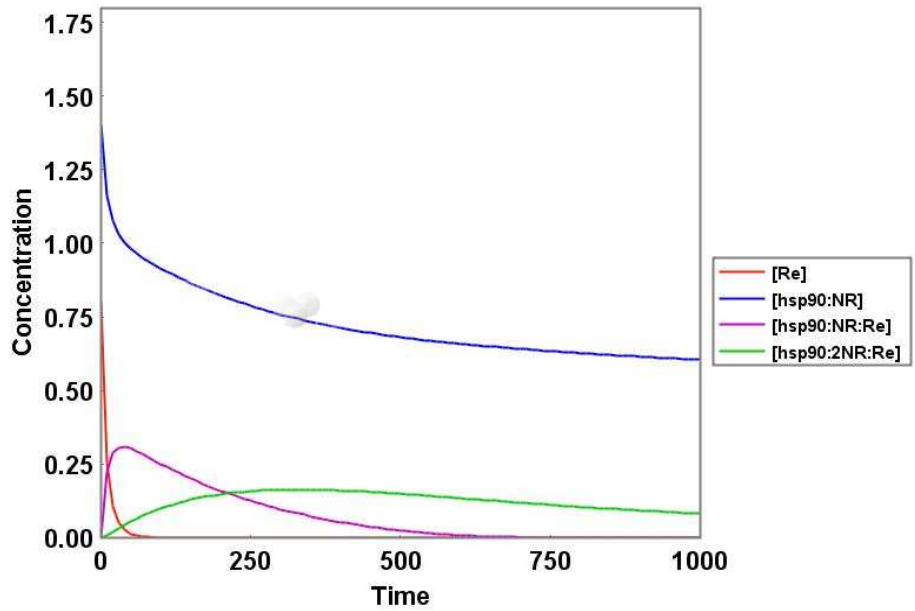}
\caption {The figure shows that the concentration of $hsp90:NR$ is in large excess. }
\label{re1} 
\end{figure}
\begin{figure}[h]            
\center
\includegraphics[width=1\textwidth]{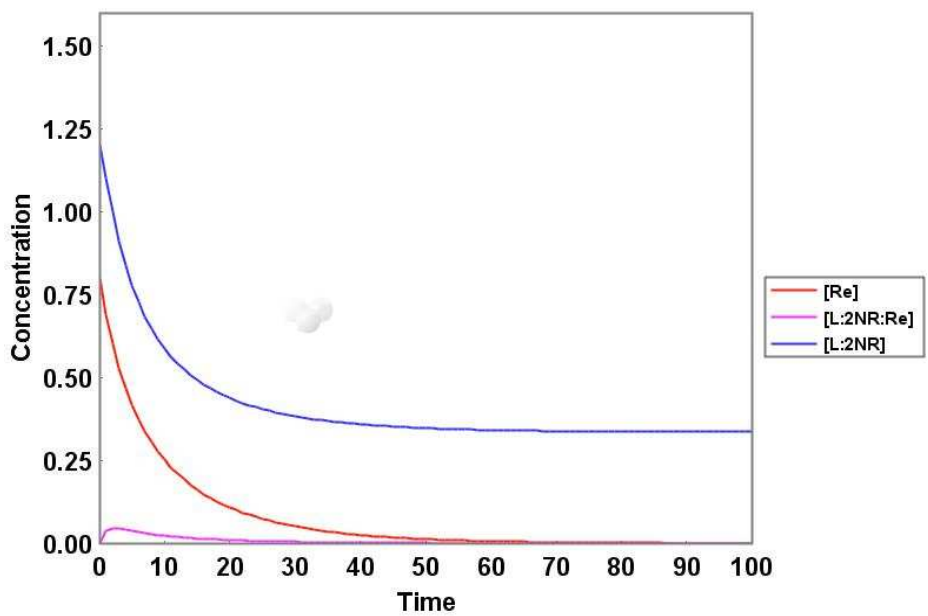}
\caption {The figure shows that the concentration of $L:2NR$ is in large excess.. }
\label{re2} 
\end{figure}
\begin{figure}[h] 
\center
\includegraphics[width=1.2\textwidth]{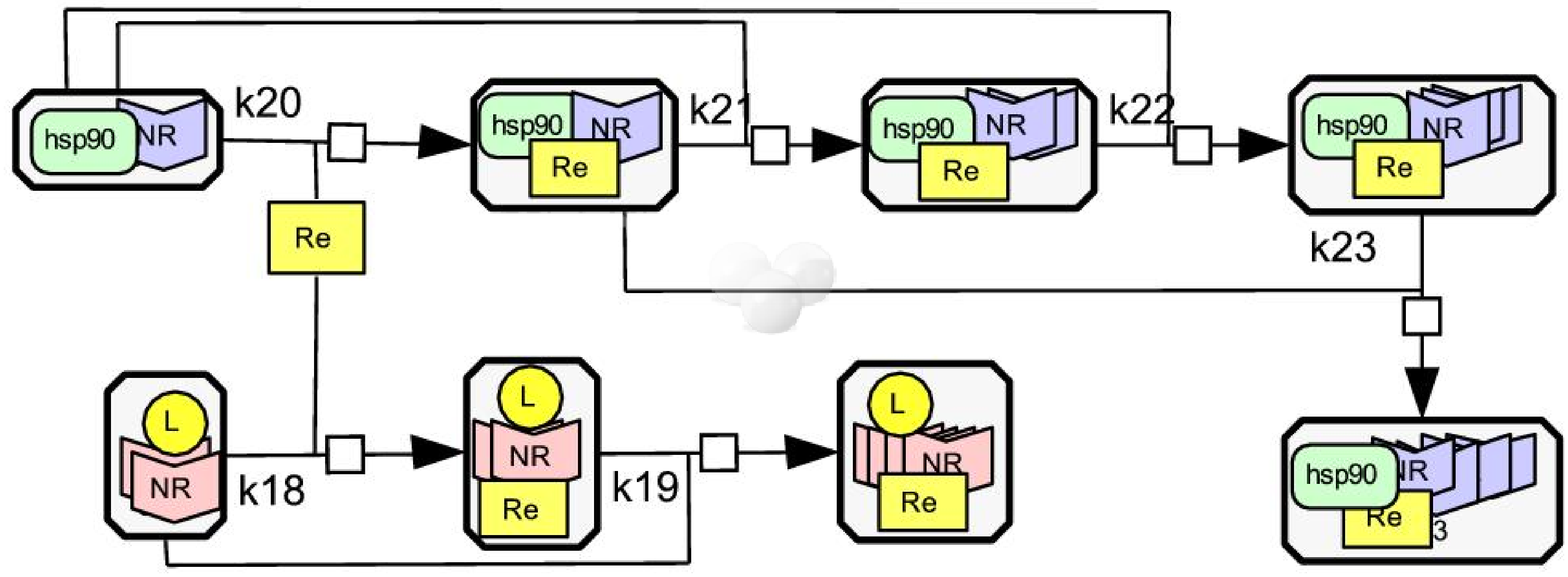}
\caption {The figure shows the original model of NRB and the model is presented by using CellDesigner. The original model was suggested in \cite{Kolodkin}}
\label{nrb1}   
\end{figure}
\begin{figure}[h] 
\center  
\includegraphics[width=1.2\textwidth]{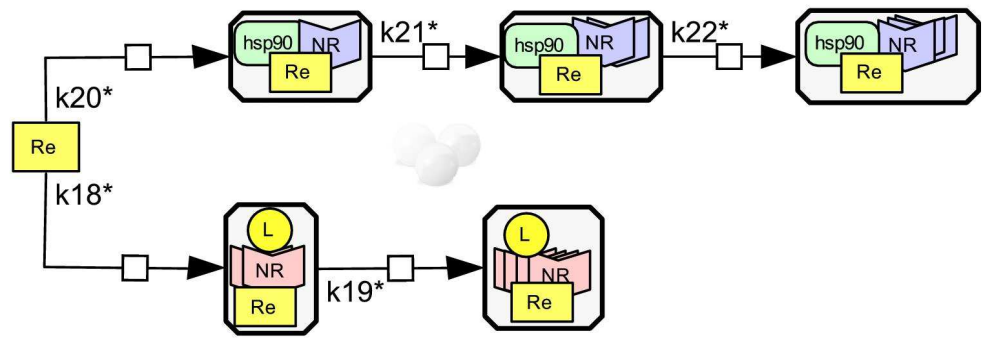}
\caption {Structure of the model of NRB after simplifying some kinetic equations based on pseudo-first order approximation, and removing the slowest reaction (reaction 23) from the model. }
\label{nrb2} 
\end{figure} 

\begin{figure}[ht]
     \begin{center}
        \subfigure[]{%
            \label{fig:third}
            \includegraphics[width=0.6\textwidth]{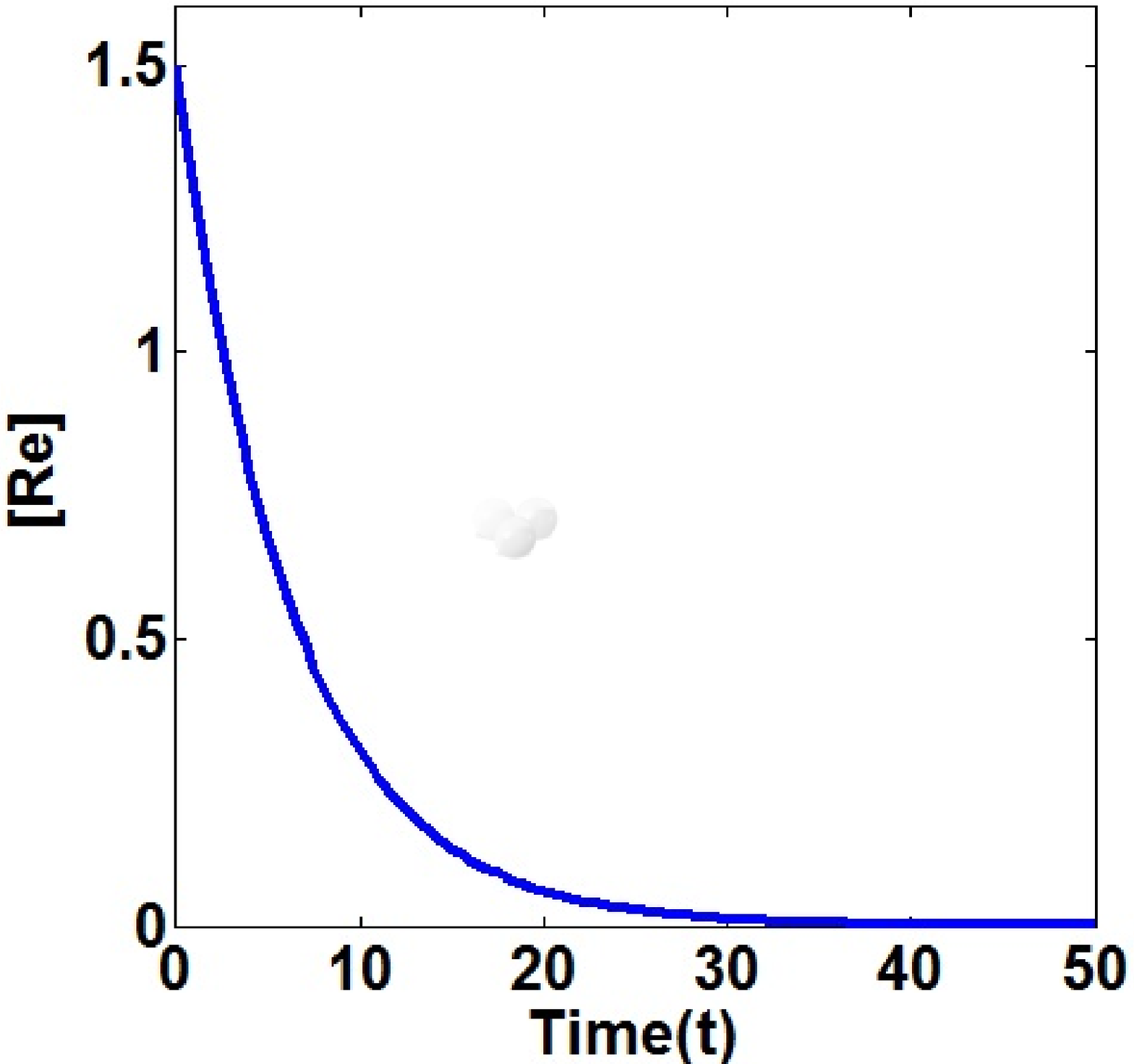}
        }%
        \subfigure[]{%
            \label{fig:fourth}  
            \includegraphics[width=0.6\textwidth]{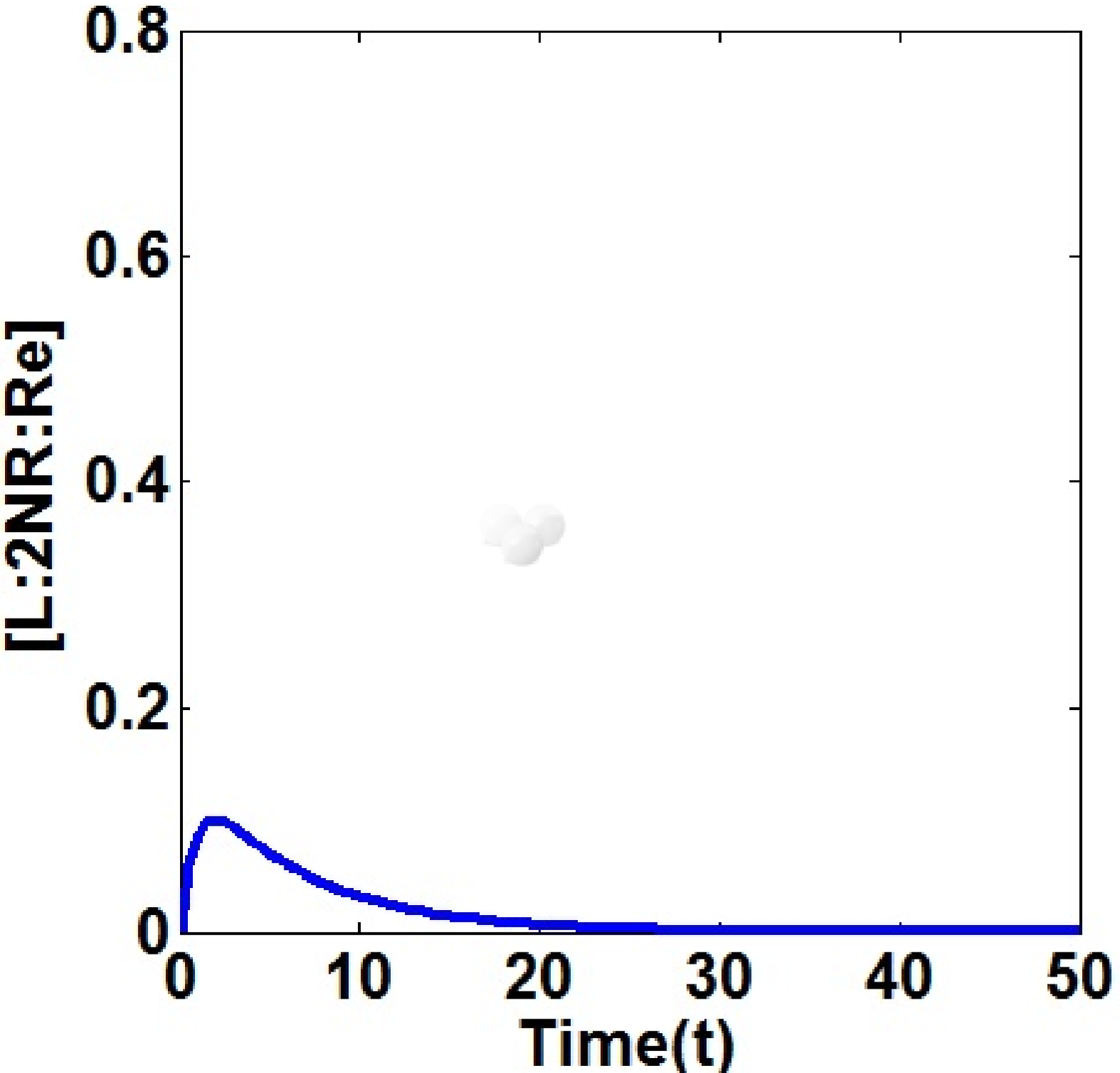}   
        } \\%
          \subfigure[]{%
            \label{fig:third}
            \includegraphics[width=0.6\textwidth]{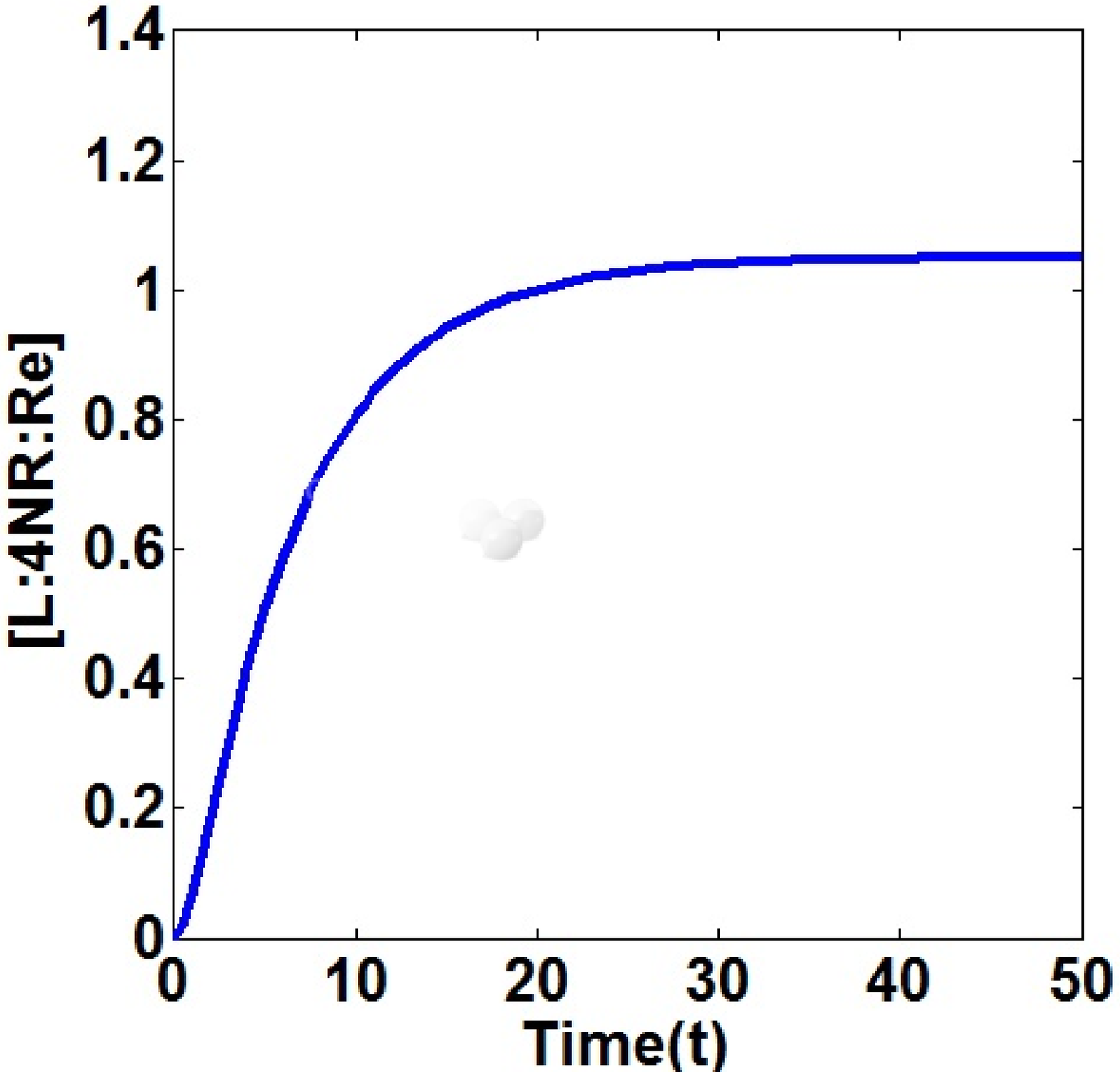}
        }%
        \subfigure[]{%
            \label{fig:fourth}
            \includegraphics[width=0.6\textwidth]{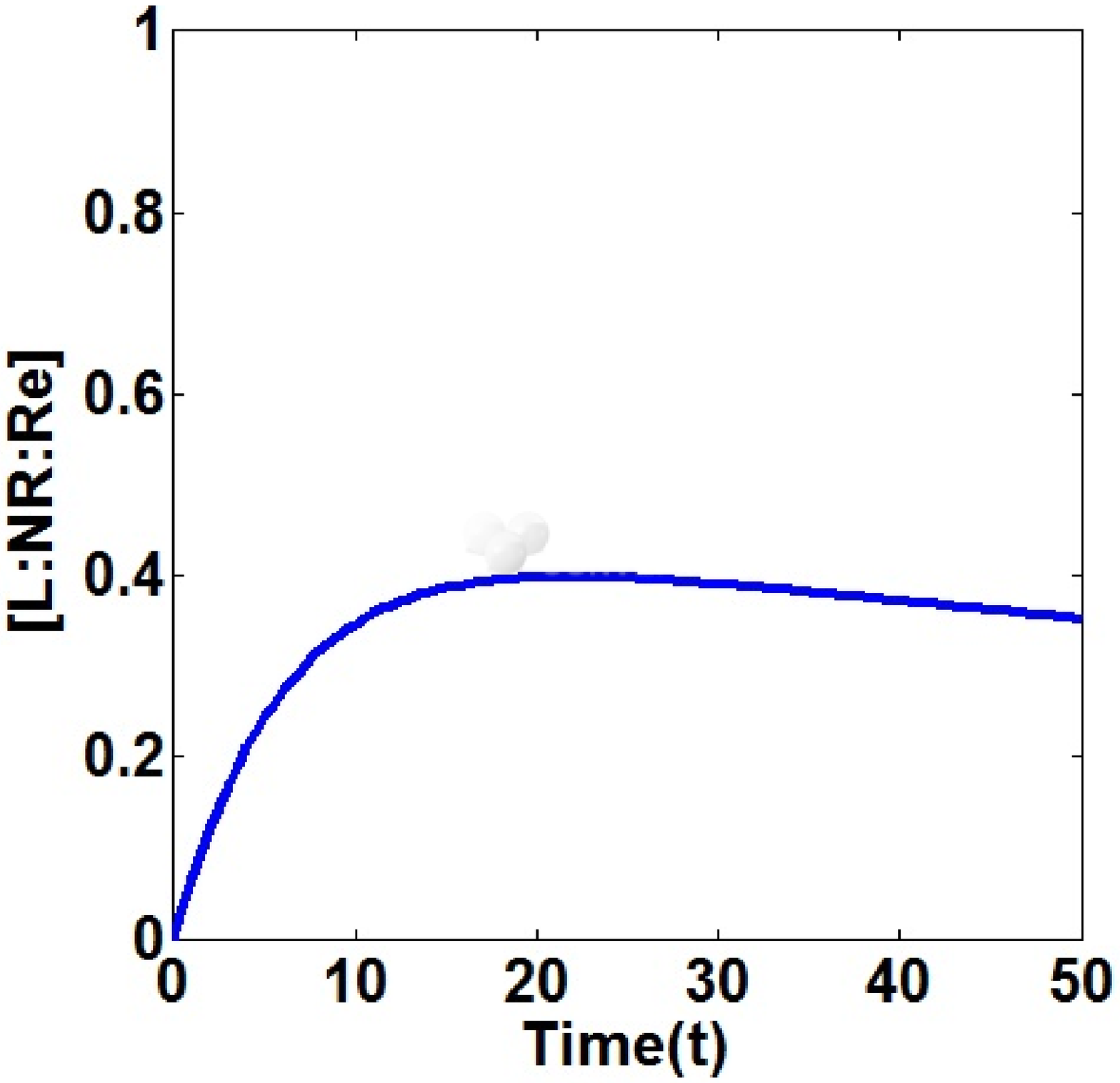}
        } \\%
          \subfigure[]{%
            \label{fig:third}
            \includegraphics[width=0.6\textwidth]{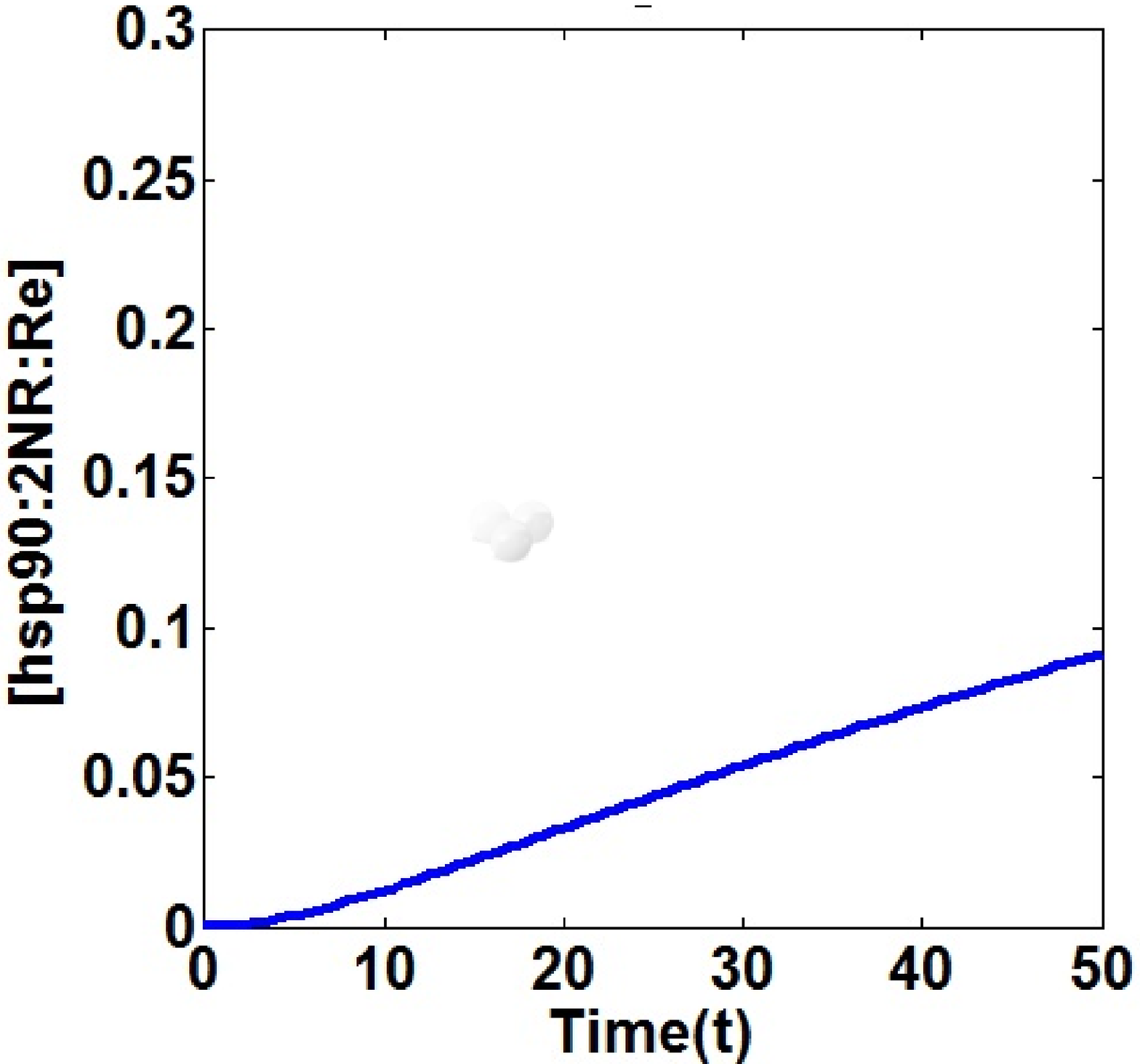}
        }%
        \subfigure[]{%
            \label{fig:fourth}
            \includegraphics[width=0.6\textwidth]{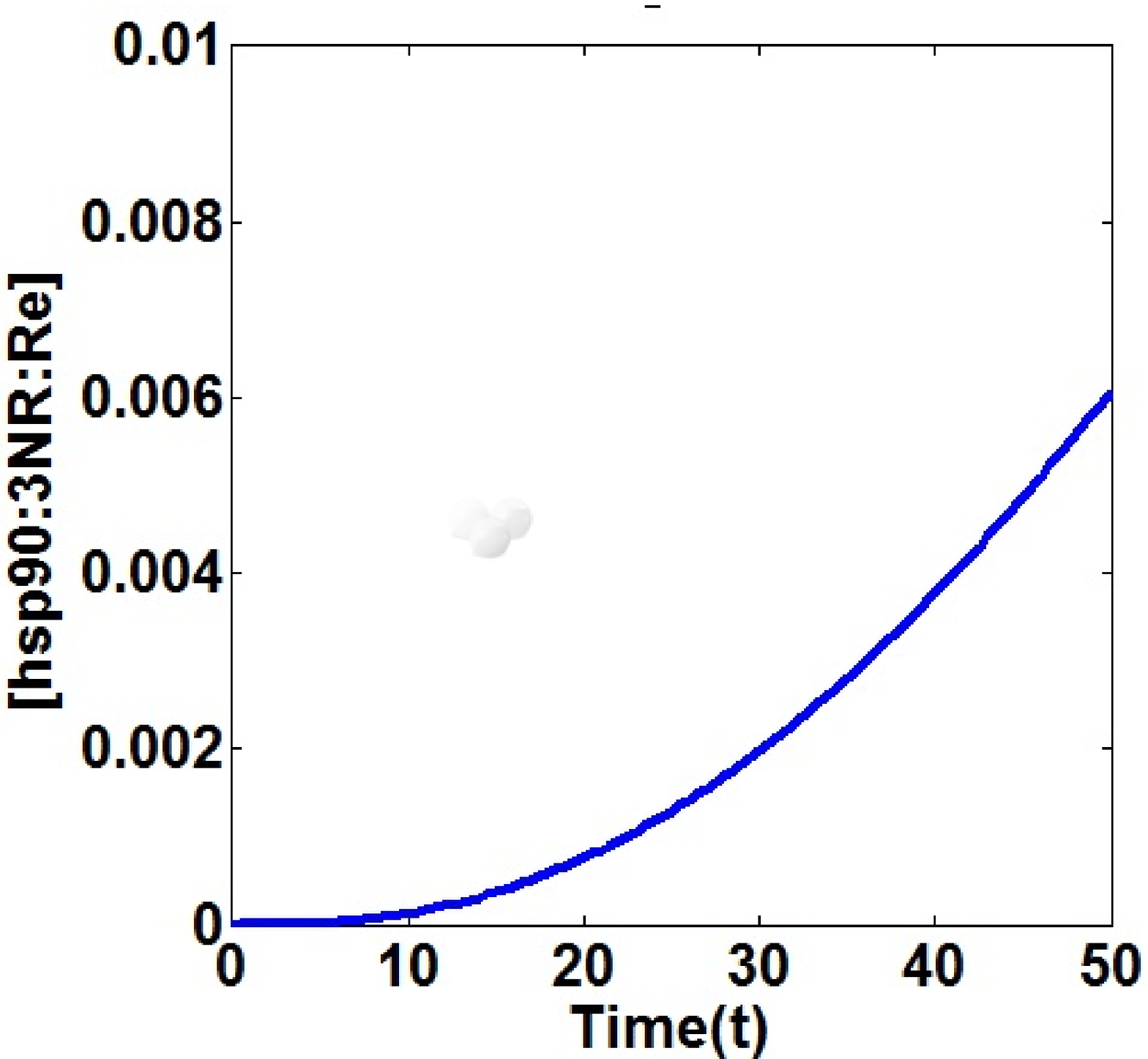}
       
        } %
    \end{center}
    \caption {The concentration of the species of the reduced model, the general analytical solution, rate constants (Table \ref{table3}) and initial concentrations (Table \ref{table4}) are used to find the concentrations.}  
   \label{A34}
\end{figure}

}

\end{document}